\documentclass[journal,comsoc]{IEEEtran}
\usepackage{amsmath,amsfonts}
\usepackage{array}
\usepackage[caption=false,font=normalsize,labelfont=sf,textfont=sf]{subfig}
\usepackage{textcomp}
\usepackage{stfloats}
\usepackage{url}
\usepackage{graphicx}
\usepackage{booktabs}
\usepackage{cite}
\usepackage{xcolor}
\usepackage{balance}
\usepackage{booktabs}
\usepackage{array}
\usepackage{tabularx}
\usepackage[table]{xcolor}
\usepackage{tikz}
\usetikzlibrary{positioning}
\definecolor{headerBlue}{HTML}{D6EAF8}  
\definecolor{lightGray}{gray}{0.95}     
\definecolor{bestGreen}{HTML}{D4EFDF}
\begin{document}

\title{Advanced AI Service Provisioning in O-RAN\\through LLM Engine Integration}
\author{Seyed~Bagher~Hashemi~Natanzi,~\IEEEmembership{Student~Member,~IEEE,}
        Pranshav~Gajjar,
        Bo~Tang,~\IEEEmembership{Senior Member,~IEEE,}
        and~Vijay~K.~Shah,~\IEEEmembership{Senior Member,~IEEE}
\thanks{S.~B.~H.~Natanzi and B.~Tang are with the Department of Electrical and Computer Engineering, Worcester Polytechnic Institute, Worcester, MA 01609 USA (e-mail: snatanzi@wpi.edu; btang1@wpi.edu).}
\thanks{P.~Gajjar and V.~K.~Shah are with the Department of Electrical and Computer Engineering, North Carolina State University, Raleigh, NC 27695 USA.}
\thanks{This work was supported by NSF under Awards CNS-2120411 and CNS-2120442.}}
\markboth{}
{Natanzi \MakeLowercase{\textit{et al.}}: Advanced AI Service Provisioning in O-RAN through LLM Engine Integration}
\maketitle

\begin{tikzpicture}[remember picture, overlay]
\node[anchor=north, yshift=-0.5cm] at (current page.north) {%
\fbox{\parbox{0.95\textwidth}{\centering\footnotesize
This work has been submitted to the IEEE for possible publication. Copyright may be transferred without notice, after which this version may no longer be accessible.
}}};
\end{tikzpicture}
\begin{abstract}
The Open Radio Access Network (O-RAN) architecture allows AI to be embedded directly into the RAN through modular xApps and rApps, yet creating these applications collecting data, training models, writing code, and deploying them safely remains slow and largely manual. Large Language Models (LLMs) offer strong reasoning and code-generation capabilities but are unsuited for the fast, deterministic inference required in real-time RAN control. We present a proof-of-concept Dual-Brain architecture that combines both strengths: an LLM-based orchestrator translates operator intents into data-collection policies and deployment code, while an automated ML engine, NeuralSmith, trains lightweight classifiers on demand via an API. We describe the architecture and provisioning workflow, share practical insights from a containerized O-RAN 5G~SA testbed, and discuss open research directions.
\end{abstract}


\section{Introduction}
\IEEEPARstart{C}{ellular} networks have evolved into software-defined, increasingly intelligent systems, with Artificial Intelligence (AI) expected at every layer of deployment and operation~\cite{9023918}.
 The Open Radio Access Network (O-RAN)~\cite{tripathi2025fundamentals} architecture is a key enabler of this vision: by disaggregating the base station and introducing intelligent controllers, O-RAN allows third-party AI applications (xApps and rApps) to be deployed directly into the RAN.
However, while \emph{deploying} AI applications is becoming easier, \emph{creating} them remains a significant bottleneck. Provisioning a new xApp involves a chain of manual handoffs across network engineers, data scientists, software engineers, and deployment engineers, often taking days to weeks far too slow for operators needing to respond to rapidly changing network conditions.

The rapid development of Large Language Models (LLMs) has led the community to ask: can LLMs serve as the AI engine inside O-RAN? The answer is nuanced. LLMs excel at understanding intents, reasoning about systems, and generating code, but they are unsuited for real-time numerical inference: their latency violates the Near-RT RIC's 10\,ms budget, their outputs are non-deterministic, and their model size makes per-xApp deployment impractical.

Recent work has begun integrating LLMs into telecommunications and O-RAN at multiple layers. Early studies explored generative AI for telecom~\cite{10.1109/MCOM.001.2300364}, while NetLLM~\cite{10.1145/3651890.3672268} adapted LLMs to networking tasks such as viewport prediction and bitrate adaptation. Closer to our domain, ORAN-Bench-13K~\cite{10975994} and ORANSight-2.0~\cite{11096935} benchmarked and fine-tuned LLMs for O-RAN knowledge, exposing the gap between general-purpose LLMs and the domain reasoning required in the RIC. At the system level, Hermes~\cite{ayed2024hermeslargelanguagemodel} proposed chained LLM agents for autonomous network modeling, LLM-xApp~\cite{WuLLMxAppAL} deployed an LLM directly inside a Near-RT RIC xApp, and recent vision work argues for agentic orchestrators replacing narrow predictive AI in 6G AI-RAN~\cite{gajjar2026agentsreplacenarrowpredictive}.

The key insight of this article is that LLMs should not \emph{replace} classical machine learning but should instead \emph{orchestrate} it. We present a Dual-Brain architecture that pairs an LLM-based Zero-Touch Orchestration Agent (ZTO-Agent) with a dedicated ML engine, NeuralSmith~\cite{NeuralSmith2024}. Unlike prior work, our architecture decouples orchestration from real-time execution: the LLM handles semantic tasks (intent parsing, code synthesis) at the Non-RT RIC layer, while purpose-built classifiers handle sub-10\,ms control decisions at the Near-RT RIC. Unlike free-form code generation, our template-constrained synthesis eliminates hallucination risk in deployed control logic. The paper describes the architecture, details the provisioning workflow, shares practical insights from a containerized O-RAN 5G~SA testbed, and discusses open research directions.

\section{O-RAN Background}
We briefly review the O-RAN architecture to establish context. Readers already familiar with O-RAN may skip to Section~III.

\subsection{Architecture Overview}
O-RAN disaggregates the base station into the O-CU, O-DU, and O-RU, connected via open interfaces~\cite{polese2023oran}. Two intelligent controllers enable closed-loop network management. The \emph{Near-RT RIC}, deployed at the network edge, operates between 10\,ms and 1\,s and hosts \emph{xApps} microservices that collect RAN data via E2SM-Key Performance Measurement (KPM) and issue control actions via E2SM-RC.
 The \emph{Non-RT RIC} operates above 1\,s and hosts \emph{rApps} for policy guidance and lifecycle management, communicating with the Near-RT RIC through the A1 interface.

  
    
    
    

Traditional AI deployments in O-RAN span supervised learning (e.g., traffic prediction), unsupervised learning (e.g., anomaly detection), and reinforcement learning (e.g., dynamic resource allocation), each typically embedded as a classifier or policy inside the Near-RT RIC. The contribution of this work is not another classifier \emph{within} the RIC, but an orchestrator at the Non-RT RIC layer that \emph{automates the creation and deployment} of such classifiers through LLM-driven intent translation, data curation, and xApp code synthesis.

As the last row highlights, the contribution of this work is not another classifier \emph{within} the RIC, but an orchestrator that \emph{automates the creation and deployment} of such classifiers.

\subsection{The Provisioning Gap}
The WG2 specification defines a conceptual AI/ML workflow but leaves implementation open. In practice, provisioning a new xApp involves a chain of manual handoffs: a network engineer extracts KPMs, a data scientist trains a model, a software engineer writes the xApp, and a deployment engineer rolls it out. Each handoff introduces delay and risk of error. This fragmented process can take hours to days far too slow for operators needing to respond to rapidly changing network conditions.

\section{The Dual-Brain Architecture}
\begin{figure*}[!t]
\centering
\includegraphics[width=0.75\textwidth]{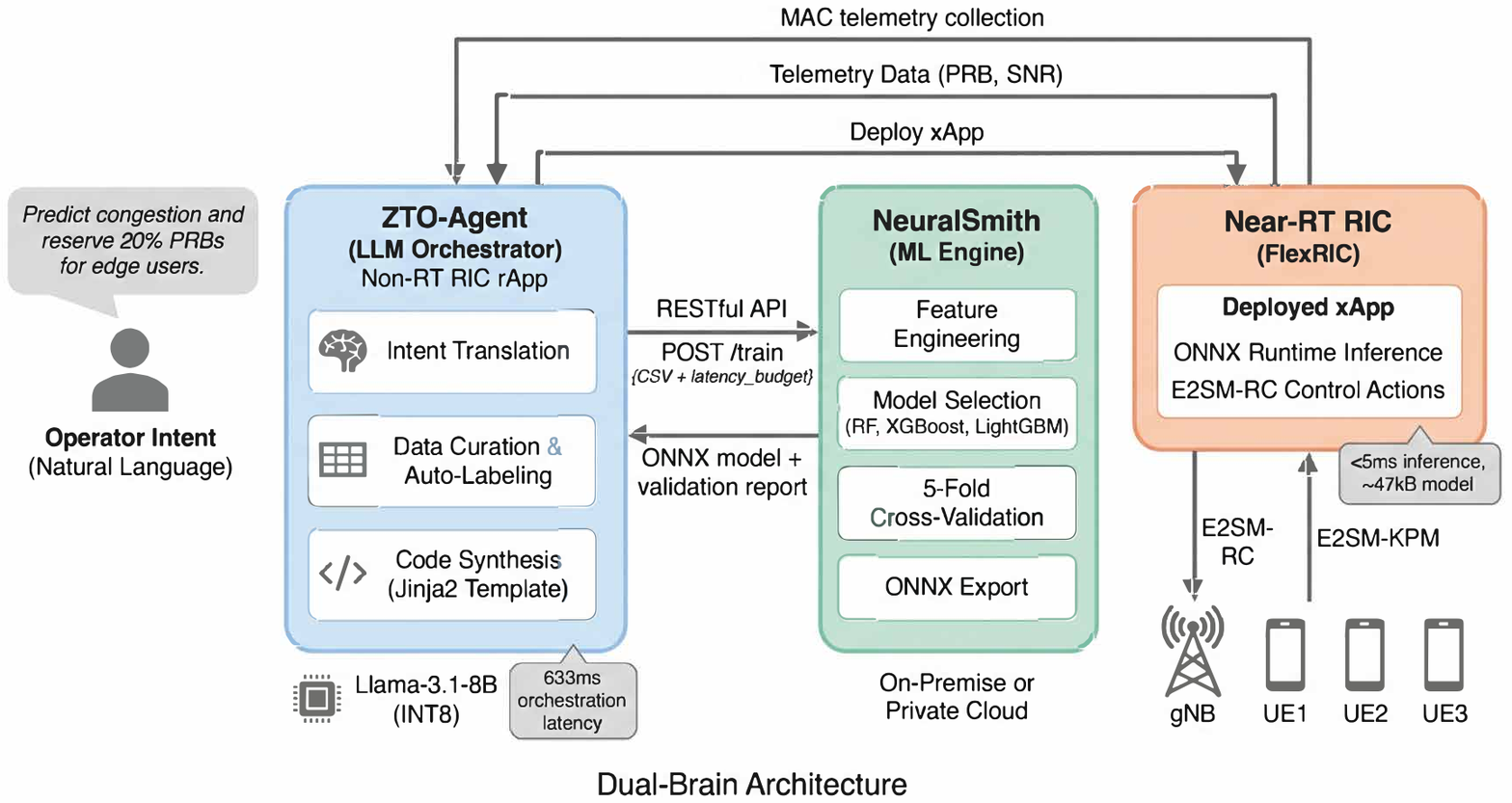}
\vspace{-35pt} %

\caption{The Dual-Brain architecture. The ZTO-Agent (LLM orchestrator, Non-RT RIC rApp) parses intents, curates data, and synthesizes xApp code. \texttt{NeuralSmith} (ML engine) trains lightweight classifiers and returns ONNX models via API. The trained model is injected into a pre-verified xApp template and deployed to the Near-RT RIC.}
\label{fig:arch}
\end{figure*}

To bridge this gap, we propose a Dual-Brain architecture (Fig.~\ref{fig:arch}) built on a simple principle: let the LLM think, and let the ML engine compute.

\subsection{Brain~1: The LLM Orchestrator (ZTO-Agent)}
The ZTO-Agent is an rApp in the Non-RT RIC powered by a foundation model (Llama-3.1-8B in our prototype, but the architecture is LLM-agnostic). It handles three tasks:

\begin{enumerate}
    \item Intent Translation: The operator submits a goal in plain language (e.g., ``\textit{predict cell-edge congestion and reserve Physical Resource Blocks (PRBs)}''). The ZTO-Agent produces a structured specification: which metrics to collect, at what granularity, how to label the data, and what model characteristics are needed.
    
    \item Data Curation: The ZTO-Agent collects MAC-layer telemetry from the gNB including per-UE Physical Resource Block (PRB) allocations and channel quality indicators and applies automatic labeling. For congestion prediction, time intervals where aggregate cell PRB utilization exceeds a saturation threshold (e.g., 80\% of total capacity) are labeled as \texttt{congested}.
    
    \item Code Synthesis: After receiving the trained model from \texttt{NeuralSmith}, the ZTO-Agent injects the ONNX artifact into a pre-verified Jinja2 xApp template containing E2SM-RC control logic and ONNX Runtime inference calls. Because the LLM only fills template variables not arbitrary code the xApp inherits the template's correctness guarantees. The container is then registered with the Near-RT RIC.
\end{enumerate}

\subsection{Brain~2: The ML Engine (\texttt{NeuralSmith})}
\texttt{NeuralSmith} is a dedicated ML engine platform that automates the full model-building workflow~\cite{NeuralSmith2024}. Upon receiving a curated dataset and a latency budget from the ZTO-Agent, it runs feature engineering, hyperparameter-optimized model selection (Random Forest, XGBoost, LightGBM, compact MLPs), 5-fold cross-validation, and exports the winning model as an ONNX artifact with a validation report.

\begin{table}[!t]
  \centering
  \caption{\textbf{Why Not Use the LLM as the Classifier?} Comparison of key properties for Near-RT RIC deployment.}
  \label{tab:llm_vs_ml}
  \small
  \renewcommand{\arraystretch}{1.25} 
  
  \begin{tabularx}{\columnwidth}{
    >{\raggedright\arraybackslash}p{2.8cm}
    !{\vrule width 0.5pt}
    >{\centering\arraybackslash}X
    !{\vrule width 0.5pt}
    >{\centering\arraybackslash}X} 
    \specialrule{0.8pt}{0.5pt}{0.5pt}
    \rowcolor{headerBlue}
    \textbf{Property} & \textbf{LLM (8B)} & \textbf{ML Classifier} \\
    \specialrule{0.8pt}{0.5pt}{0.5pt}
    Inference Latency & 100--500\,ms & \cellcolor{bestGreen}\textbf{$<$5\,ms} \\
    Determinism & No & \cellcolor{bestGreen}\textbf{Yes} \\
    Model Size & 4--16\,GB & \cellcolor{bestGreen}\textbf{10--500\,kB} \\
    Near-RT RIC Ready & No & \cellcolor{bestGreen}\textbf{Yes} \\
    \specialrule{0.8pt}{0.5pt}{0.5pt}
  \end{tabularx}
\end{table}

Table~\ref{tab:llm_vs_ml} illustrates why this dedicated engine is necessary. An 8B-parameter LLM requires hundreds of milliseconds per inference far exceeding the Near-RT RIC's 10\,ms budget and produces non-deterministic outputs unsuitable for safety-critical control. \texttt{NeuralSmith} produces classifiers that are three to four orders of magnitude smaller, deterministic, and fast. While LLMs are ideal for high-level reasoning, using them for sub-10\,ms inference is computationally disproportionate. We therefore delegate real-time execution to specialized engines like \texttt{NeuralSmith}, ensuring both deterministic performance and computational economy at the edge.

\subsection{Deployment Flexibility}
A critical concern for operators is data sovereignty. \texttt{NeuralSmith} supports flexible deployment: fully on-premise (telemetry never leaves the operator's perimeter) or in a private cloud, depending on contractual requirements. Both the ZTO-Agent and \texttt{NeuralSmith} can run on the same compute cluster, keeping the entire pipeline within the operator's trust boundary.

\section{Provisioning Workflow}
The complete lifecycle proceeds in four phases (Fig.~\ref{fig:workflow}):

\begin{figure*}[!t]
\centering
\includegraphics[width=1\linewidth]{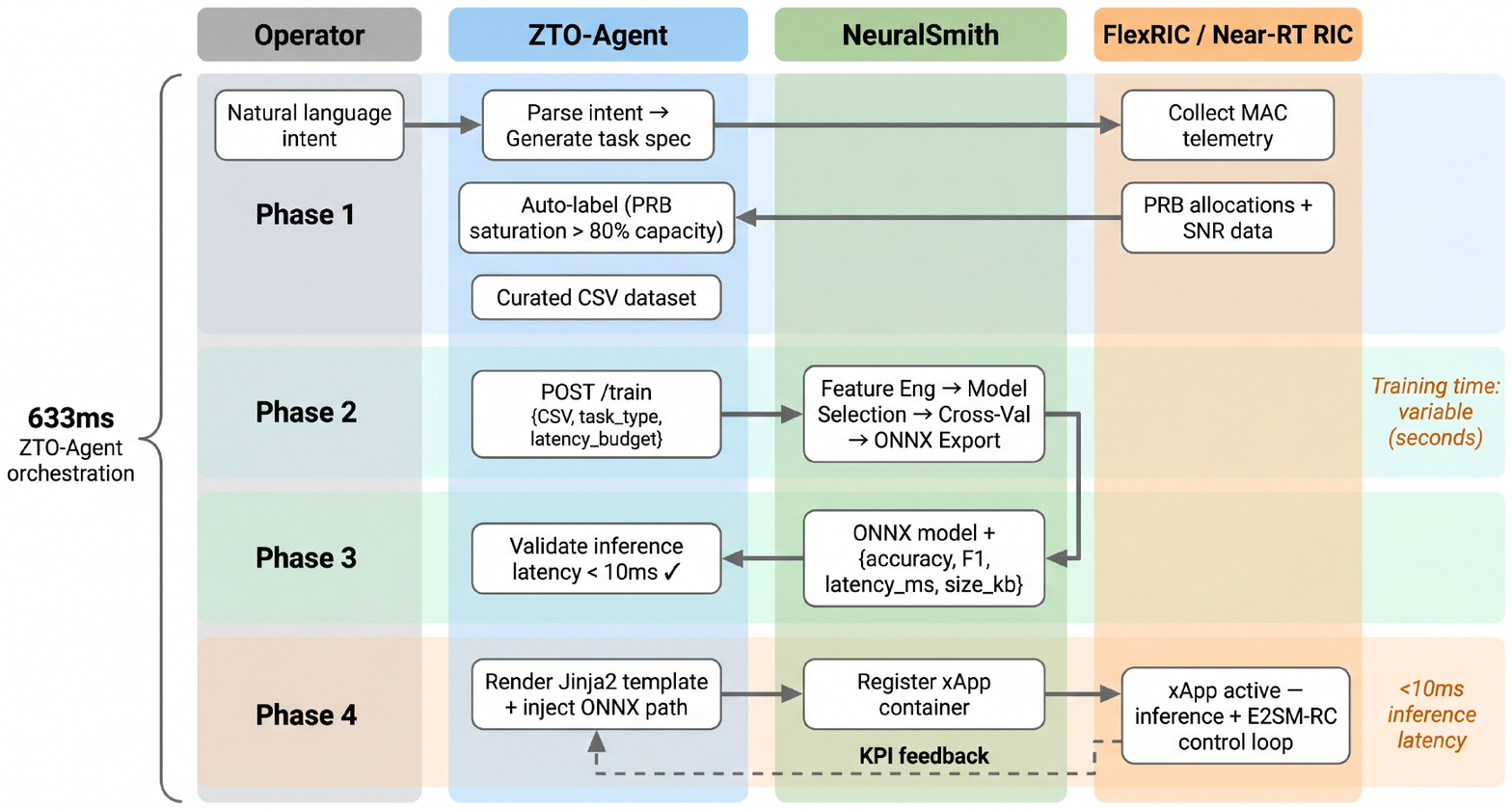}

\vspace{-35pt} %

\caption{Four-phase provisioning workflow: (1)~intent and telemetry subscription, (2)~curated data to \texttt{NeuralSmith} API, (3)~trained ONNX model returned, (4)~xApp template rendered and deployed. \textbf{Note:} While the diagram reflects an initial cold-start orchestration latency of 633\,ms, the system achieves an optimized steady-state (warm) latency of 384\,ms in operational deployments.}
\label{fig:workflow}
\end{figure*}

\noindent\textit{Phase~1 (Intent$\to$Data).} The ZTO-Agent parses the operator's natural-language goal, collects MAC-layer telemetry from the gNB (per-UE PRB allocations and SNR), and auto-labels intervals where aggregate cell resource utilization exceeds saturation.

\noindent\textit{Phase~2 (Data$\to$NeuralSmith).} The labeled dataset is passed to NeuralSmith's API with a task type and latency budget; NeuralSmith trains, validates, and exports an ONNX classifier, and the ZTO-Agent resubmits with a tighter constraint if the latency budget is exceeded.

\noindent\textit{Phase~3 (Model$\to$Live xApp).} The ZTO-Agent renders the Jinja2 template with the model path and action parameters, containerizes it, and registers it with FlexRIC; the xApp begins executing within the Near-RT RIC's control loop.

End-to-end orchestration overhead is 384\,ms under steady-state conditions on an Ollama-served Llama-3.1-8B endpoint, well within the Non-RT RIC's 1\,s budget, and can be further reduced on dedicated GPU infrastructure.

\section{Implementation and Practical Insights}
Rather than presenting exhaustive experimental results, we share the practical insights gained from building and operating the Dual-Brain architecture on a containerized O-RAN 5G~SA testbed. Our setup uses Docker-containerized OpenAirInterface (OAI) components~\cite{oai2024} with the rfsimulator, three simulated UEs with \texttt{iperf3} traffic generation, and FlexRIC~\cite{flexric2023} providing E2 interface connectivity. Telemetry is extracted from the gNB's MAC-layer scheduler logs, which provide per-UE PRB allocation, BLER, and SNR measurements at sub-second granularity. Although we evaluate the system on this minimal topology to isolate core behaviors, containerizing the ZTO-Agent and \texttt{NeuralSmith} as microservices ensures the architecture inherently scales to multi-cell, multi-vendor deployments.

\subsection{What Worked Well}
\textit{Separation of concerns.} Early in our development, we attempted to use the LLM for both orchestration and inference, prompting it to classify congestion states from raw telemetry values. The results were poor: latency was unacceptable, outputs were inconsistent across identical inputs, and the model occasionally hallucinated congestion states that did not exist. Separating the LLM (thinking) from the ML engine (computing) resolved all three issues immediately.

\textit{Template-based code synthesis.} By restricting the LLM to populating pre-verified Jinja2 templates rather than generating arbitrary xApp code, we eliminated the most dangerous failure mode of LLM-based code generation: hallucinated control logic reaching the live network. Every xApp produced by our pipeline inherits the template's tested correctness. This approach directly addresses the industry's requirement for a robust AI Agent Harness; by confining the LLM's output to predefined parameter spaces, our architecture enforces a structural guardrail that ensures autonomous code generation adheres to standardized signaling flows and safety protocols.

\textit{Sufficiency of rfsimulator for orchestration benchmarking.} Because the Dual-Brain pipeline operates entirely above the physical layer, consuming MAC-layer telemetry and issuing E2 control actions, the software-simulated RF channel faithfully exercises every interface the system uses. The deterministic channel also makes experiments reproducible, which is valuable for iterative development.

\textit{Latency-constrained training.} Specifying an inference-latency budget in NeuralSmith's training request ensures that the returned model is always compatible with the Near-RT RIC's timing constraints. Without this, the automated pipeline would risk producing accurate but too-slow models.

\subsection{What Was Challenging}
\textit{Auto-labeling threshold sensitivity.} The ZTO-Agent's automatic labeling, which marks intervals where aggregate cell PRB utilization exceeds 80\% capacity as congested, worked well for our scenario, but the choice of threshold significantly affects model quality. In practice, adaptive or multi-threshold labeling strategies may be needed for more nuanced tasks.

\textit{Intent ambiguity.} Operator intents like ``protect cell-edge users'' are ambiguous: protect from what? Congestion? Interference? Handover failures? The ZTO-Agent sometimes misinterpreted vague intents. Structured intent templates with clarifying prompts improved reliability significantly.

\begin{figure}[!t]
\centering
\includegraphics[width=1\columnwidth]{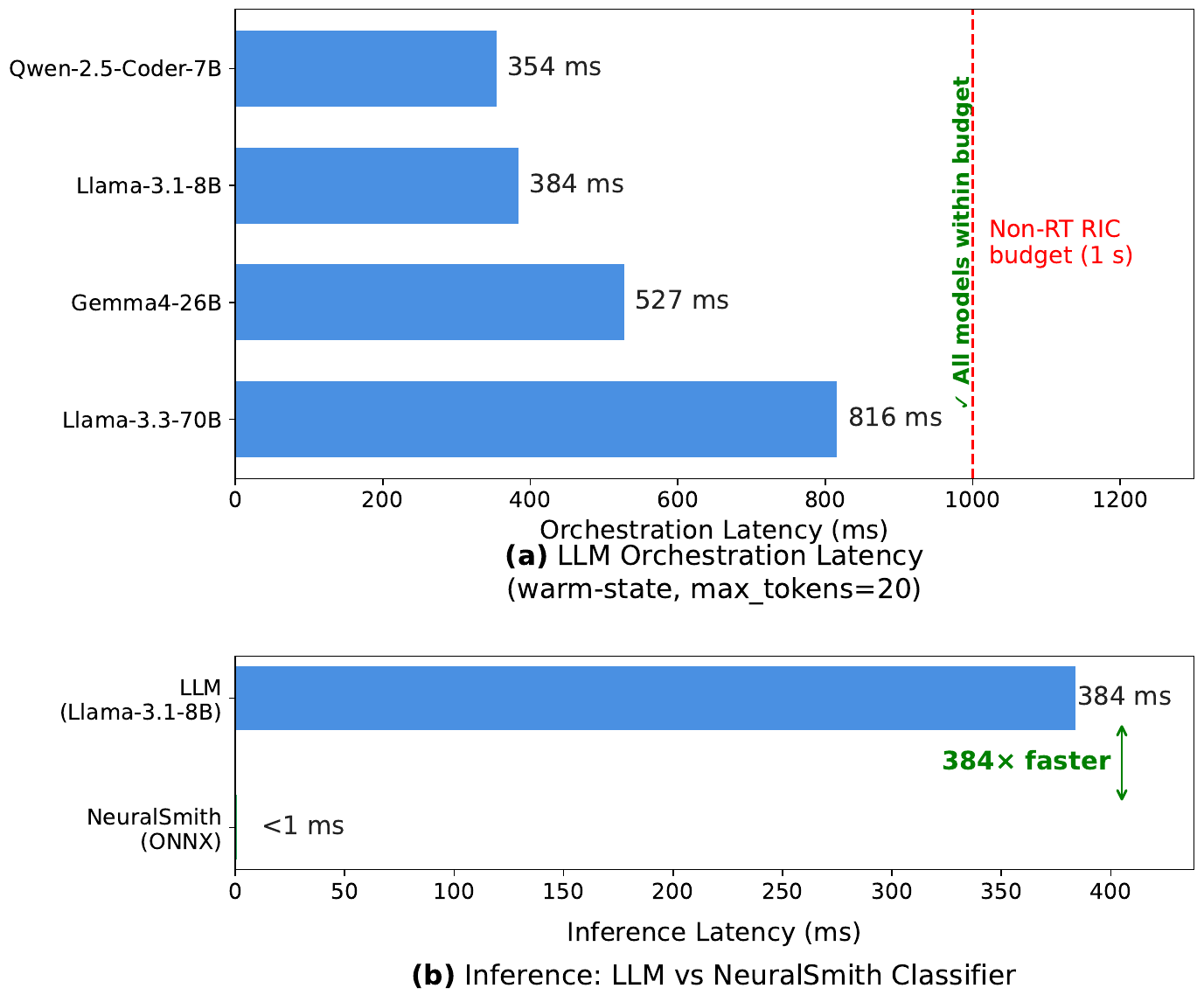}
\caption{ZTO-Agent orchestration latency comparison across four foundation models, and the resulting inference-latency gap between LLM and NeuralSmith classifier. (a) All four models complete orchestration well within the Non-RT RIC's 1\,s budget, with Llama-3.1-8B at 384\,ms and Llama-3.3-70B at 816\,ms. (b) Once trained, the NeuralSmith ONNX classifier delivers inference at sub-millisecond latency, roughly 384$\times$ faster than direct LLM inference and well within the Near-RT RIC's 10\,ms control loop budget.}

\label{fig:benchmark}
\end{figure}

\textit{Deployment-dependent LLM latency.} The 384\,ms steady-state orchestration time with Llama-3.1-8B served via Ollama is well within the Non-RT RIC's timing budget. However, smaller models or CPU-only deployments may struggle to meet the 1\,s threshold for complex intents requiring multi-step reasoning. Repeating the pipeline with three additional foundation models (Qwen-2.5-Coder-7B, Gemma4-26B, and Llama-3.3-70B) yielded identical xApps in each case, identical in terms of generated Jinja2 template parameters and resulting E2SM-RC action configuration. As compared in Fig.~\ref{fig:benchmark}, the higher latency of larger models stemmed from unoptimized local serving rather than structural limits, confirming the architecture remains LLM-agnostic when properly hosted.

\subsection{Congestion Management Demonstration}
\begin{figure}[!t]
\centering
\includegraphics[width=0.8\columnwidth]{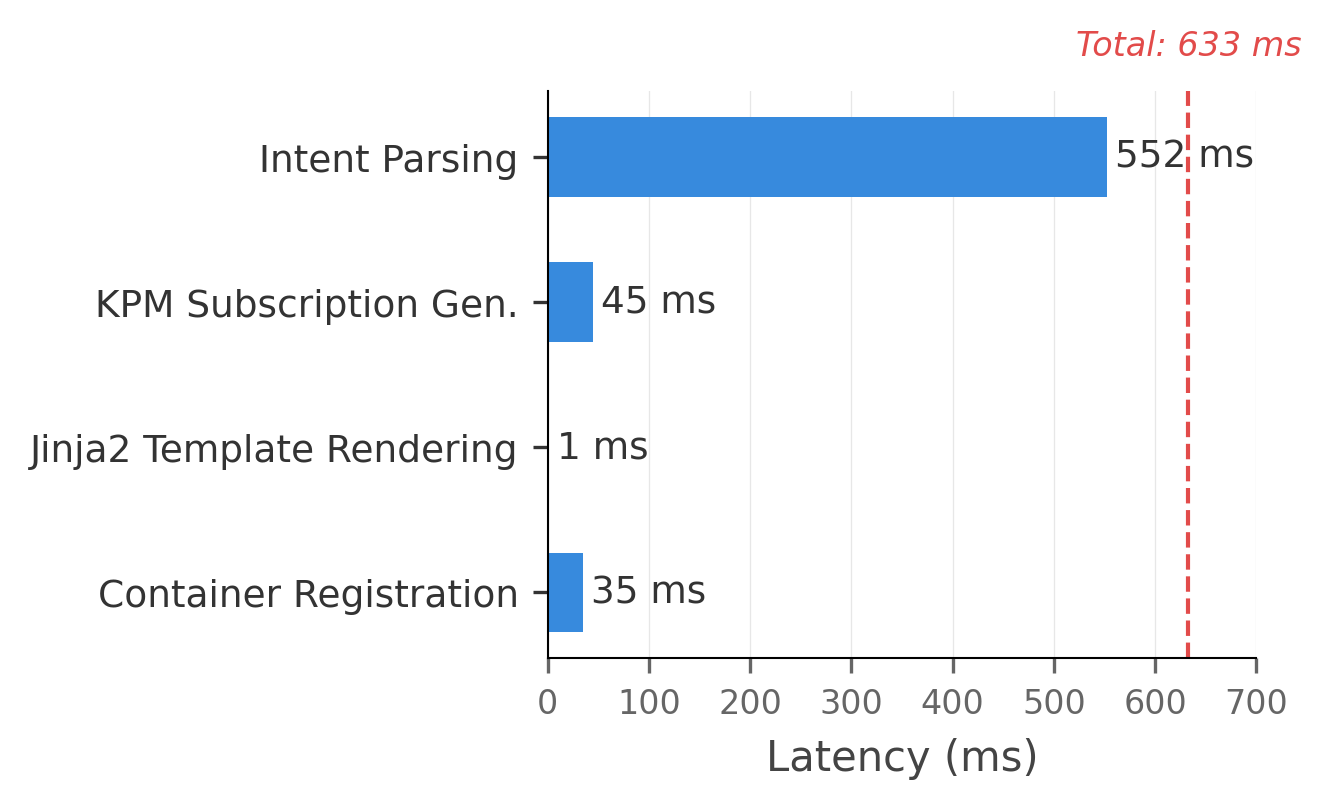}
\caption{ZTO-Agent (Llama-3.1-8B via Ollama) orchestration latency is dominated by intent parsing (552\,ms); all other phases (KPM subscription generation, Jinja2 template rendering, and container registration) complete in under 82\,ms combined. Total latency is 633\,ms cold-start and 384\,ms warm-state, well within the Non-RT RIC's 1\,s timing budget.}

\label{fig:latency}
\end{figure}

To demonstrate the proposed Dual-Brain architecture pipeline, we deploy a predictive congestion-management xApp.
 In this scenario, two center UEs generate bursty UDP traffic via \texttt{iperf3} at 20\,Mbps each in six on-off cycles over 20~minutes, while a third cell-edge UE maintains a low-rate background session. The operator inputs the intent: ``\textit{predict congestion and reserve 20\% PRBs for edge users.}'' The ZTO-Agent collects MAC-layer telemetry (per-UE PRB allocations and SNR), labels intervals where aggregate cell PRB utilization exceeds 80\% capacity as congested, and dispatches the curated dataset to \texttt{NeuralSmith}. \texttt{NeuralSmith} returns a LightGBM classifier (ONNX, 49\,kB) achieving $97.7\%$ accuracy (F1: 0.975) with sub-millisecond inference latency on CPU well within the Near-RT RIC's 10\,ms budget. The ZTO-Agent then renders the xApp template with the trained model and deploys it to FlexRIC. Although the rfsimulator's deterministic AWGN channel limits raw feature diversity (SNR and Block Error Rate (BLER) are near-constant), temporal feature engineering via sliding-window aggregation over PRB allocations enables \texttt{NeuralSmith} to learn congestion onset patterns with high fidelity. Fig.~\ref{fig:latency} breaks down the orchestration latency by phase, confirming that intent parsing dominates the pipeline while downstream synthesis and deployment add minimal overhead. Table~\ref{tab:results} summarizes the complete set of measured pipeline metrics, including orchestration latency across the four foundation models we tested and the NeuralSmith classifier performance.

To assess the value of automated ML over hand-coded heuristics, we compared NeuralSmith's LightGBM output against a simple threshold rule that flags congestion whenever instantaneous PRB utilization exceeds 80\%. The threshold rule achieves 84\% accuracy on the same test set but reacts only to saturation that has already occurred. In contrast, the LightGBM classifier, using temporal features such as PRB-utilization mean and slope over a sliding window, anticipates congestion onset 2--3 measurement intervals earlier, demonstrating the value of automated feature engineering in the Dual-Brain pipeline.

\begin{table}[!t]
  \centering
  \caption{Dual-Brain Pipeline Evaluation Results}
  \label{tab:results}
  \small
  \renewcommand{\arraystretch}{1.25}
  \begin{tabularx}{\columnwidth}{
    >{\raggedright\arraybackslash}X 
    !{\vrule width 0.5pt} 
    >{\raggedright\arraybackslash}p{3.5cm}}
    \specialrule{0.8pt}{0.5pt}{0.5pt}
    \rowcolor{headerBlue}
    \textbf{Metric} & \textbf{Value} \\
    \specialrule{0.8pt}{0.5pt}{0.5pt}
    \multicolumn{2}{l}{\cellcolor{lightGray}\textit{ZTO-Agent Orchestration}} \\
    Llama-3.1-8B Latency         & 384\,ms (warm) \\
    Qwen-2.5-Coder-7B Latency   & 354\,ms (warm) \\
    Gemma4-26B Latency           & 527\,ms (warm) \\
    Llama-3.3-70B Latency        & 816\,ms (warm) \\
    xApp Output Consistency      & \cellcolor{bestGreen}\textbf{Identical} \\
    \specialrule{0.5pt}{0.5pt}{0.5pt}
    \multicolumn{2}{l}{\cellcolor{lightGray}\textit{\texttt{NeuralSmith} Classifier}} \\
    Winning Algorithm            & LightGBM \\
    Accuracy                     & 97.7\% \\
    F1-Score (macro)             & 0.975 \\
    ONNX Inference Latency       & $<$1\,ms \\
    ONNX Model Size              & 49\,kB \\
    Training Features            & \texttt{mean\_prb, std\_prb, min\_prb, slope\_prb} \\
    \specialrule{0.8pt}{0.5pt}{0.5pt}
  \end{tabularx}
\end{table}

\section{Research Directions}
The Dual-Brain architecture opens several avenues for future research, which we group into four themes that we believe are most consequential for scaling LLM-orchestrated O-RAN provisioning to production deployments.

\textit{From Simulation to Real Deployments.}
Our prototype operates over a software-simulated RF channel, which faithfully exercises the orchestration interfaces but masks the variability of real wireless environments. Bridging this gap raises two coupled questions: how to make NeuralSmith-trained classifiers robust to channel fading, interference, and hardware impairments through techniques such as domain randomization and online adaptation; and how to overcome the metadata fragmentation that currently prevents models trained at one site from generalizing to another. The community would benefit substantially from standardized reference datasets, effectively an MNIST for wireless, that capture both telemetry and environmental context, enabling reproducible sim-to-real transfer and cross-vendor portability of autonomously provisioned xApps.

\textit{Safety, Trust, and Lifecycle Management.}
Autonomously generated xApps introduce safety concerns that classical hand-crafted control logic does not. Three directions appear particularly pressing. First, autonomously synthesized xApps must be subjected to systematic correctness, robustness, and adversarial testing before deployment, extending prior O-RAN AI testing frameworks~\cite{10077111} into the LLM-orchestrated regime. Second, when multiple autonomously provisioned xApps coexist, their control actions may conflict; for example, a throughput-boosting xApp competing with an energy-saving xApp for the same PRB pool. Pre-deployment conflict detection, possibly through static analysis of the rendered templates or guard-band enforcement on shared parameters, is an important safeguard. Third, operators and regulators need to understand what each automatically generated xApp does and why; integrating explainability reports that surface the features the classifier relied upon and the decisions the LLM made during synthesis is essential for production adoption.

\textit{Closed-Loop, Confidence-Aware Orchestration.}
The current pipeline is trigger-based: the operator submits an intent, an xApp is built and deployed, and the system waits for the next intent. Two extensions would transform this into a continuous learning loop. The ZTO-Agent could monitor live xApp KPIs and re-trigger NeuralSmith upon detecting performance drift, enabling self-healing AI services without operator intervention. In parallel, the orchestration layer itself would benefit from calibrated uncertainty estimation. LLMs at the heart of the ZTO-Agent can produce overconfident outputs under ambiguous network conditions, and recent work on conformal prediction for LLM-based autonomous control~\cite{farzaneh2026iexpresseddifferentintent} offers a principled way to attach reliability guarantees to intent parsing and auto-labeling decisions. Low-confidence cases could then be flagged for human review, bridging the gap between full autonomy and operator trust.

\textit{Efficient and Portable LLM Backends.}
Two practical concerns determine how broadly the Dual-Brain architecture can be deployed. The first is computational cost: while an 8B-parameter LLM runs comfortably on GPU-equipped Non-RT RIC infrastructure, smaller cell sites and edge deployments will need lighter backends. Research on telecom-specialized small language models, parameter-efficient fine-tuning, and speculative decoding tuned to the structured nature of orchestration prompts could substantially reduce this barrier. The second is portability: generalizing xApp templates across vendor-specific E2 implementations requires deeper standardization of xApp packaging and E2SM profile negotiation, so that the same intent reliably produces an interoperable xApp regardless of the underlying RIC platform.

\section{Conclusion}
The process of creating and deploying AI services in O-RAN remains overwhelmingly manual. This article has presented a Dual-Brain architecture that combines the orchestration capabilities of LLMs with the numerical precision of dedicated ML engines to automate the entire provisioning lifecycle. The LLM-based orchestrator handles intent translation, data curation, and code synthesis; while the ML engine, NeuralSmith, trains lightweight, deterministic classifiers via a simple API. Our experience building the system on a containerized O-RAN testbed shows that the approach is practical, achieving 384\,ms steady-state orchestration latency while meeting the Near-RT RIC's sub-10\,ms inference constraint. While our prototype demonstrates the pipeline end-to-end on a single downstream task, validation in over-the-air deployments and across additional use cases remains important future work. The architecture is LLM-agnostic, preserves data sovereignty, and offers a reproducible path toward automated AI service delivery in 6G networks.

\balance

\section*{Acknowledgment}
This work was supported by the National Science Foundation under Awards CNS-2120411 and CNS-2120442. NeuralSmith is developed by All Things Intelligence, co-founded by V. K. Shah.

\begin{IEEEbiographynophoto}{Seyed Bagher Hashemi Natanzi}
(Milad Natanzi) is a Ph.D.\ candidate in the Electrical and Computer Engineering Department at Worcester Polytechnic Institute (WPI). His research focuses on AI/ML integration for O-RAN architectures and security in 5G and 6G wireless networks.
\end{IEEEbiographynophoto}

\begin{IEEEbiographynophoto}{Pranshav Gajjar}
is a graduate research assistant in the Department of Electrical and Computer Engineering at North Carolina State University. His research interests include LLM applications for telecommunications, O-RAN benchmarking, and automated ML engineering.
\end{IEEEbiographynophoto}

 \begin{IEEEbiographynophoto}{Bo Tang}
   is an Associate Professor in the Department of Electrical and Computer Engineering at Worcester Polytechnic Institute. His research interests lie in artificial intelligence, AI security, and their applications in next-generation wireless networks. He is currently an Associate Editor for IEEE Transactions on Neural Networks and Learning Systems.
 \end{IEEEbiographynophoto}

\begin{IEEEbiographynophoto}{Vijay K. Shah}
is an Assistant Professor in the Department of Electrical and Computer Engineering at North Carolina State University, where he directs the NextG Wireless Lab. He is the Outreach Director for AERPAW and Co-Founder of All Things Intelligence, the creator of \texttt{NeuralSmith}. He co-authored \emph{Fundamentals of O-RAN} (IEEE Press/Wiley). His research focuses on 5G/6G systems, O-RAN, and wireless security.
\end{IEEEbiographynophoto}

\vfill
\renewcommand{\baselinestretch}{0.95}\selectfont
\bibliographystyle{IEEEtran}
\bibliography{bib/ref}

\end{document}